\documentclass[twocolumn,english,showpacs,preprintnumbers,amsmath,amssymb,floatfix]{revtex4}

\usepackage[T1]{fontenc}
\usepackage{float}
\usepackage{graphicx}



\usepackage{babel}

\begin{document}

\title{On the Efficiency of Data Representation on the Modeling and
  Characterization of Complex Networks}

\author{Carlos A. Ruggiero}

\author{Odemir M. Bruno}

\author{Gonzalo Travieso}

\author{Luciano da Fontoura Costa}

\affiliation{Institute of Physics at S\~ao Carlos, University of S\~ao
  Paulo, PO Box 369, S\~ao Carlos, S\~ao Paulo, 13560-970 Brazil}

\begin{abstract}
  Specific choices about how to represent complex networks can have a
  substantial effect on the execution time required for the respective
  construction and analysis of those structures.  In this work we
  report a comparison of the effects of representing complex networks
  statically as matrices or dynamically as spase structures.  Three
  theoretical models of complex networks are considered: two types of
  Erd\H{o}s-R\'enyi as well as the Barab\'asi-Albert model.  We
  investigated the effect of the different representations with
  respect to the construction and measurement of several topological
  properties (i.e. degree, clustering coefficient, shortest path
  length, and betweeness centrality).  We found that different forms
  of representation generally have a substantial effect on the
  execution time, with the sparse representation frequently resulting
  in remarkably superior performance.
\end{abstract}


\maketitle

\section{Introduction}

As a consequence of the intrinsic difficulties in achieving analytical
approaches for the characterization and modelling of natural systems,
a great deal of such investigations has to rely on computational
methods.  The typical case involves the application of numerical
methods in order to solve differential equations
(e.g.~\cite{press92:_numer_recip_in_c,naka:2005}), which is the most
frequent situation found in practice in Physics. Frequently, such
problems involve large amounts of data, as well as data which are very
large. Given the importance of effectively tackling these problems, a
lot of attention and efforts have been invested in developing,
implementing and applying numerical methods which are fast and
accurate (e.g.~\cite{press92:_numer_recip_in_c}).  Indeed, such
efforts give rise to the important area of
\emph{Computational Physics}.

A peculiar situation in computational physics is found in complex
networks research~\cite{Barabasi:survey, Dorogovtsev02,
Newman:2003:survey}, a new multidisciplinary area of physics which has
undergone an impressive development along the last decade. Here, the
investigations rely not mainly on numerical solution of differential
equations, but on intensive handling of matrices as well as
combinatorial or spectral methods as required for calculation of
measurements~\cite{Costa_surv:2007} such as shortest paths, betweeness
centrality, and spectra of graphs. Though presenting such a
distinctive nature, computational approaches to complex networks also
aim at achieving precision and speed. The latter demand often becomes
particularly critical as a consequence of the large size of several
complex networks of current interest, such as the
Internet~\cite{Faloutsos1999}, protein-protein
interaction~\cite{Jeong01:Nature}, and social
interactions~\cite{Wasserman94}, to name but a few cases.

Indeed, the effective approach to most of the remaining challenges in
complex networks research is immediately related to the ability to
effectively represent and process large structures. This can be done
in the two following ways: (i) development of more effective
algorithms; and (ii) careful and efficient respective implementation
of those algorithms. While much attention has been placed recently on
(i), the final performance will ultimately depend critically on the
implementation, making step (ii) particularly critical for achieving
good results. The current work focuses on important practical
implementational aspects related to the use of sparse or full
representation of graphs. As such the present article constitutes one
of the few works investigating the effect of such important practical
choices on the resulting efficiency of the implementation of a set of
crucially important operations typically performed in complex networks
research, including network generation as well as the estimation of
important topological properties such as the degree, clustering
coefficient, shortest path length, and betweeness centrality.

This article starts by describing the computational tasks to be
performed, namely the estimation of several topological features of
the networks, and follows by presenting the adopted network models and
the two types of representations of networks to be compared.  The work
concludes by presenting and disucussing the computational efficiency
of these two representations as obtained through computational
simulations.

\section{The methods chosen for the evaluation}

It is henceforth assumed that all networks are undirected and
unweighted.  Full representations of the networks are performed in
terms of the respective \emph{adjacency matrices} $K$, such that the
presence of an edge between nodes $i$ and $j$ imply $K(i,j)=K(j,i)=1$,
with $K(i,j)=K(j,i)=0$ being otherwise imposed. The total number of
nodes and edges in the networks are respectively abbreviated as $N$
and $E$. A set of four representative methods/measurements of complex
networks have been selected in order to investigate the effect of
implementational parameters and choices on the respective performance:
degree, clustering coefficient, shortest path and betweeness
centrality. Each of these methods are briefly revised in the
following.

\emph{Degree:} The degree of a node $i$ corresponds to the number of
links attached to it. It can be calculated by adding all entries in
column $i$ of the adjacency matrix. The degree is an intrinsically
local measurement, in the sense of taking into account only the links
directly attached to the node. Usually, the degree is calculated for
all the nodes of a given network.

\emph{Clustering Coefficient:} The clustering coefficient is also a
local measurement, specific to each node $i$. However, it also
consider the interconnectivity between the neighbors of that node.  In
the case of full representation in terms of the adjacency matrix, the
calculation of this measurements requires access to all the columns
corresponding to each of the neighbors of node $i$.

\emph{Shortest Path Identification:} Given two nodes $i$ and $j$, the
shortest topological path between them corresponds to the path which
has the smallest number of edges. Note that it is possible to have two
or more distinct shortest paths of the same size.

\emph{Betweeness Centrality:} The betweeness centrality is a property
associated to a given node or edge. In both cases, it refers to the
number of shortest paths, considering all pairs of nodes in the
networks, which pass through the given node or edge. The calculation
of the betweeness centrality requires the determination of the
shortest paths for every pair of distinct nodes.

\section{Network models}

In the following we use three network models.  Two models due to
Erd\H{o}s and R\'{e}nyi (ER) and the scale free model of Barab\'asi
and Albert \cite{Barabasi97} (BA). The first model, denoted ER
(probability), connects each pair of vertices with a fixed probability
$p$.  The average degree in this model is $p(N-1)$, where $N$ is the
number of vertices in the network.  The second model, denoted ER
(edges), uses a fixed number $E$ of edges, and connects each edge to a
randomly chosen pair of nodes.  The average degree of the network is
$2E/N$.  These two models have similar statistical properties, but are
included here due to their different behavior during network
construction: as the first model must consider all pairs of nodes, it
is computationally intensive for large networks; the second model has
construction time proportional to the number of edges, and is
therefore faster for sparse networks.

The Barab\'asi-Albert networks are constructed starting with a small
number of vertices, and adding vertices one by one, each new vertex
being connected to $m$ existing vertices, chosen using a linear
preferential attachment rule, vertices with higher degrees having
higher probabilities of being chosen.  The resulting average degree is
given by $2m$.

\section{Full or sparse representation of the networks}

Two main representations were used: adjacency matrix and adjacency
lists~\cite{cormen09:_introd_to_algor}.  Adjacency matrix is a dense
representation, in the sense that all possible edges in the network
are explicitly included, with a value used to indicate the presence of
each edge, and another value used otherwise.  Adjacency lists are
sparse, as only the edges present in the network are incorporated.
The adjacency matrix is usually implemented as a static structures,
like basic arrays used in most computer languages.  On other hand,
adjacency lists are implemented as dynamic structures and require
pointers (a memory position pointing to another one).

For the elements of the adjacency matrix, we consider five
possibilities, depending on the C language data type used for each
element: double precision (\texttt{double}, 64~bits) and single
precision (\texttt{float}, 32~bits) floating point number, integer
numbers (\texttt{int}, 32~bits), boolean values (which can assume only
true and false values, 8~bits) and bits.  This last element
representations does not have a corresponding type in C, and was
implemented using an \texttt{int} value to store 32~elements of the
matrix, with bit manipulation operations used to access the individual
bit values.

In the adjacency lists representation, a list is maintained for each
vertex, with the numbers of the vertices that are neighbors to it.
This representation uses an integer value for each neighbor and an
overhead for list administration.  Nevertheless, it spares memory
space for sparse networks.

\section{Results and discussion}

We study the execution time needed for the generation of the network
and for the computation of the following network measurements: average
degree, clustering coefficient, all-pairs distances and betweenness
centrality. The effect of the various network representations is
evaluated as a function of the network size and average degree.

\subsection{Network generation  times for different network sizes}
\label{sec:netsize}

We first consider the effect of network size on the execution time
needed for the generation of the networks, using different network
representations.

\subsubsection{ER (probability) model}

\begin{figure}
  \begin{center}
    \includegraphics[width=0.9\columnwidth]{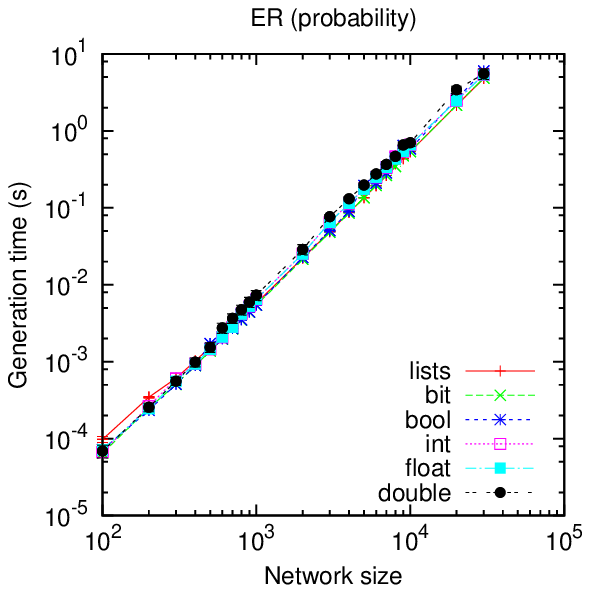} (a)
    \includegraphics[width=0.9\columnwidth]{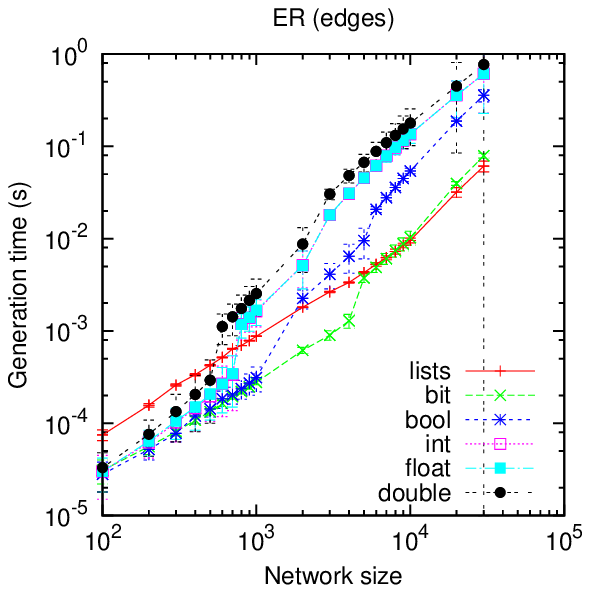} (b)
    \includegraphics[width=0.9\columnwidth]{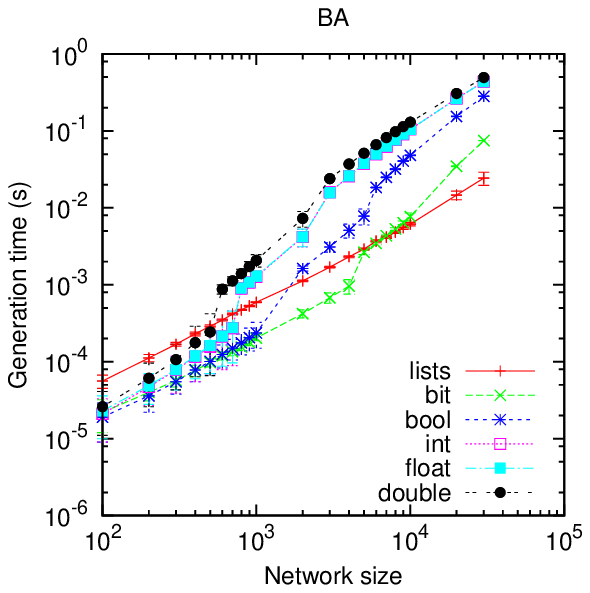} (c)
  \end{center}
  \caption{Time taken to generate networks of different sizes (number
    of nodes) for the Erd\H{o}s-R\'enyi model with fixed probability
    (a), the Erd\H{o}s-R\'enyi model with fixed number of edges (b)
    and the Barab\'asi-Albert model (c), using various graph
    representations.}  \label{fig:gen}
\end{figure}

We start by investigating how the generation of ER networks is
affected by the choice of different graph representations.  The
execution times required to produce ER networks of several sizes, and
average degree 10, are shown in Figure~\ref{fig:gen}(a).

Generally speaking, the different types of graph representation
clearly had little effect on the execution time.

The reason why the execution times resulted similar is that most of
the computational effort is invested in considering all pairs of nodes
to be connected with constant probability.

\subsubsection{ER (edges) model}

Figure~\ref{fig:gen}(b) depicts the execution times obtained for
generation of ER networks with edges for different networks sizes and
average degree equal to 10.

Unlike the results obtained previously, now the adoption of different
types of graph representations has a marked effect on the respective
execution times.  In particular, the improvements allowed by the more
memory-effective representations (bit and list) are now evident.
Interestingly, a sharp change of execution times in the matrix cases
is observed at about $N = 1000$.  This abrupt increases occurs when
the capacity of the cache of the microcomputer is exceeded by larger
sizes of graphs.  Though the list representation is initially slower
than the matrix cases, it becomes faster and faster with the increase
of $N$.

The substantial differences now observed between the execution times
obtained for the diverse representations are a consequence of the fact
that the smaller time required for the choice of the pairs to be
connected implies that the intrinsic access time for each type of
representation becomes more pronounced.

\subsubsection{BA model}

The generation times obtained for BA networks with average degree 10
are shown in Figure~\ref{fig:gen}(c).  As with ER, the list and bit
representations provide the fastest execution times for large values
of $N$.  Along the region where the cache is large enough to cope with
the graph size, the bits representation is the fastest option, but
with the increase of $N$ the list implementation becomes progressively
more effective.  This is a consequence of the fact that the
computational cost with a list representation is linear with $N$,
while cost with the matrix representation increases with $N^2$.  As
before, the matrix-based implementations tend to be slower.

\subsection{Computation time of some measurements for different
  network sizes}

We turn now our attention to the effect of network size on the
computation time of some network measurements, for the various network
representations.

\subsubsection{Average degree}

\begin{figure*}
  \begin{center}
    \includegraphics[width=0.9\columnwidth]{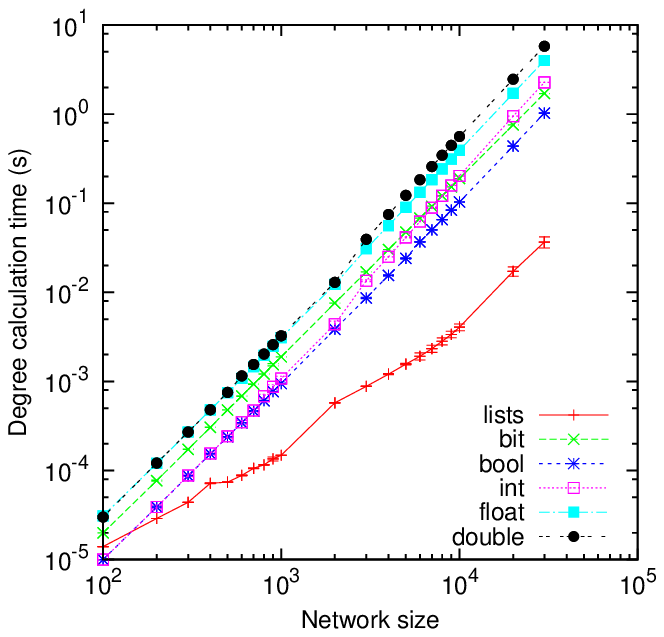} (a)
    \includegraphics[width=0.9\columnwidth]{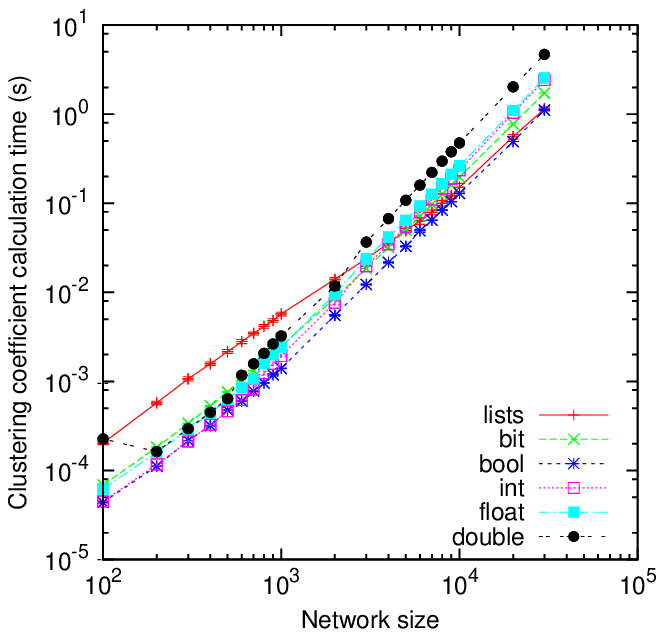} (b)
    \includegraphics[width=0.9\columnwidth]{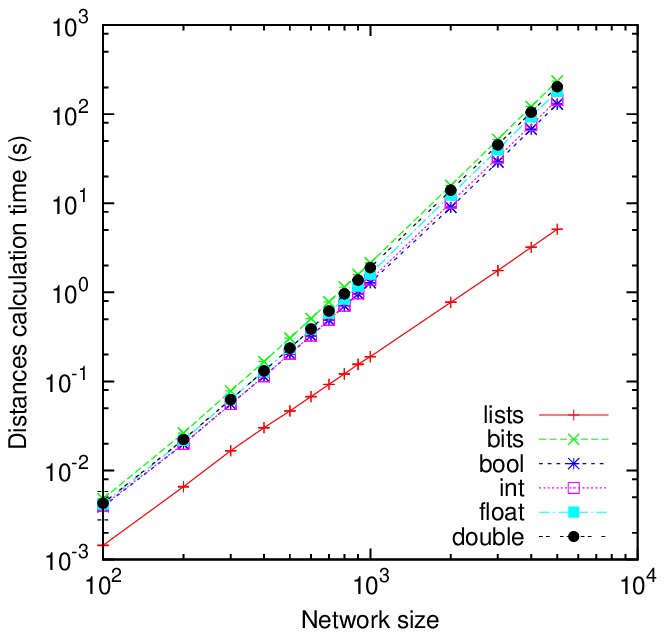} (c)
    \includegraphics[width=0.9\columnwidth]{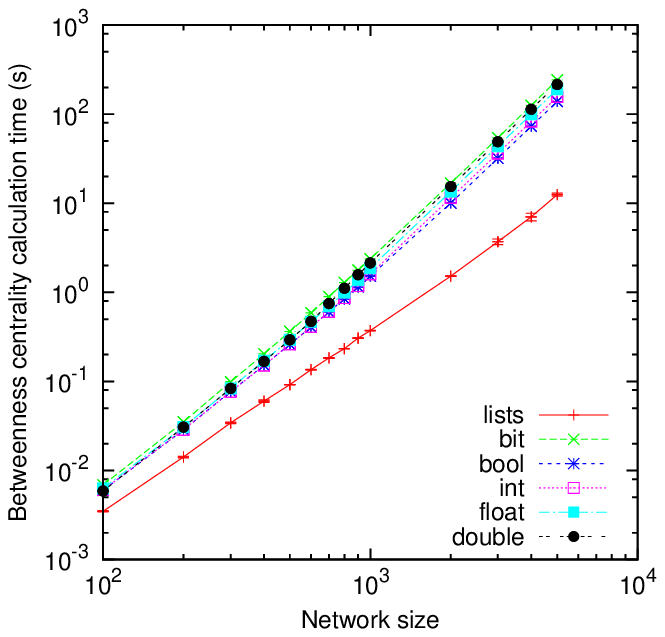} (d)
  \end{center}
  \caption{Computation time for some network measurements [average
    degree (a), clutering coefficient (b), average distances (c) and
    node betweenness centrality (d)] as a function of network size for
    BA networks of average degree~$10$.  } \label{fig:calc}
\end{figure*}

We now turn our attention to the effect of different graph
representations on the execution time required for the calculation of
some of the principal measurements of the topology of the graphs.  We
start by investigating the execution times required for determination
of the average degree.  Figure~\ref{fig:calc}(a) depicts the
obtained results for BA networks with varying sizes and average degree
equal to 10.

It is clear that the list implementation allows a dramatic reduction
of the execution times for most network sizes.  The other
implementations required similar execution times and, as could be
expected, the double representation implied the longest execution
times.

\subsubsection{Clustering coefficient}

Figure~\ref{fig:calc}(b) shows the execution times obtained for the
calculation of the clustering coefficient of BA networks of several
sizes and average degree 10.  As a consequence of the fact that this
measurement demands more computations than the average degree, the
execution times resulted larger than those in
Figure~\ref{fig:calc}(a).  Interestingly, the several tested
representations led to similar execution times, with the list and bool
implementations providing particularly good efficiency for large
values of $N$.

\subsubsection{All-pairs distances}

Figure~\ref{fig:calc}(c) presents the execution times obtained while
calculating the average shortest distance lengths for several BA
networks with average degree equal to 10.  Similarly to the betweeness
centrality, this measurement also requires intensive computations.

The results are similar to those obtained for the betweeness, but the
relative improvement allowed by the lists implementation is still
larger now.

\subsubsection{Betweenness centrality}

We also investigated how the time required for the calculation of the
betweeness centrality varied with the several adopted implementations.
Figure~\ref{fig:calc}(d) shows the obtained results for BA networks
of several sizes and average degree equal to 10.

The substantially more complex nature of this measurement has been
clearly reflected in the larger execution times.  While little
differences can be noticed for most implementations, the list
representation allowed, again, substantially faster execution times,
representing the fast option for all values of $N$.  Indeed, the
relative improvement obtained with lists clearly seems to increase
with the network size.  This implies that the use of lists becomes
critical for allowing calculation of betweeness in particularly large
networks.

\subsection{Network generation  times for different average degrees}

So far we have probed how the execution time varies with the network
size for a fixed average degree (equal to 10).  Now, we proceed to
investigate how the speed is influenced by different values of average
degree.  This will allow us to get insights about the generality of
the previously observed trends.  In principle, it could be expected
that the larger the average degree of a network, the smaller would be
the benefits provided by the lists, because the matrices would become
less sparse.  Therefore, special attention is henceforth focused on
this potential effect.

\subsubsection{Network generation time for the BA model}

\begin{figure}
  \begin{center}
    \includegraphics[width=\columnwidth]{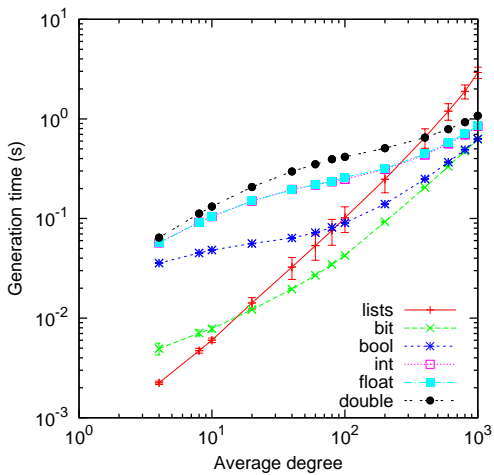}
  \end{center}
  \caption{Network generation time (BA model with $N=10000$) as a
    function of average degree.} 
  \label{fig:time_gen}
\end{figure}

Figure~\ref{fig:time_gen} shows the execution times, in terms of the
average degree, obtained for generating BA networks while using the
several representations.  The network size is henceforth fixed at $N =
10000$.

The results are evident and confirm that the use of lists guarantees
higher speed up to about average degree 100, decreasing steeply
thereafter.  Particularly interesting is the behavior of the bits
implementation, which overtakes the lists from average degree 20.
This fact suggests that the execution time seems to be strongly
affected by the memory which is demanded by each implementation.  With
the increase of the average degree, the matrix implementations become
progressively more effective, while the bits, and particularly the
list, implementations loose their effectiveness. 

\subsection{Computation time of some measurements for different
  average degrees}

Now we consider the effect of the average degree in the computation of
some network measurements for different graph representations.

\subsubsection{Average degree}

\begin{figure*}
  \begin{center}
    \includegraphics[width=0.9\columnwidth]{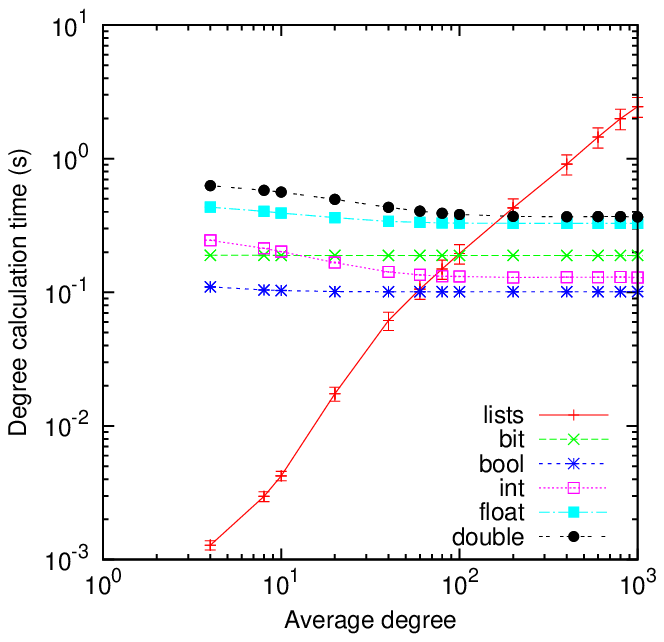} (a)
    \includegraphics[width=0.9\columnwidth]{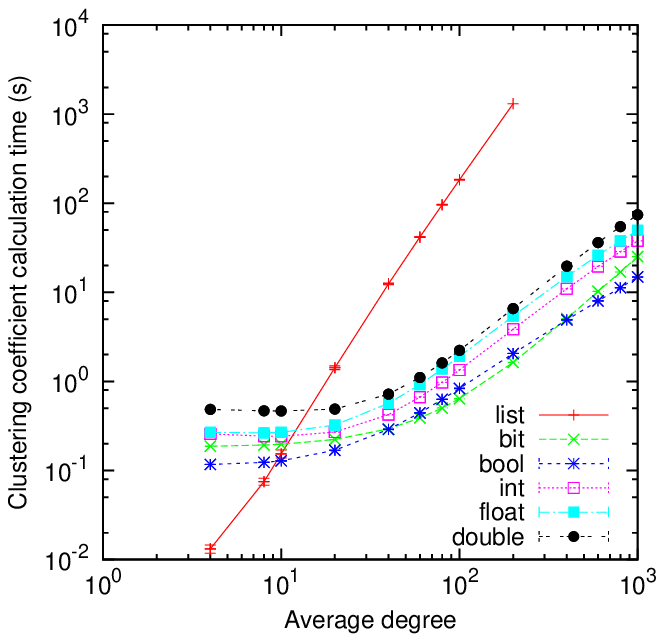} (b)
    \includegraphics[width=0.9\columnwidth]{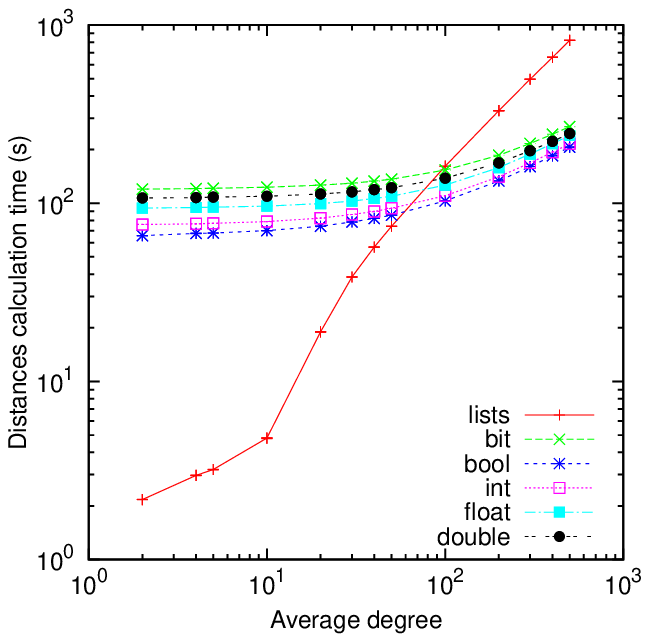} (c)
    \includegraphics[width=0.9\columnwidth]{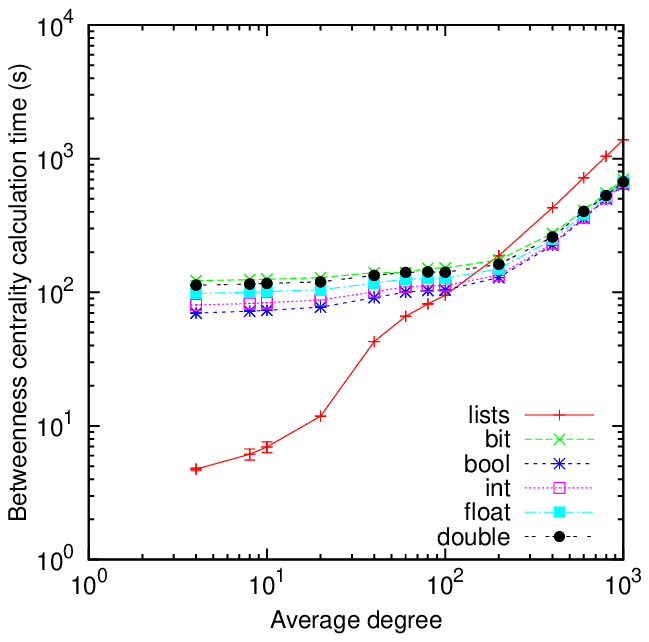} (d)
  \end{center}
  \caption{Computation time for some network measurements [average
    degree (a), clutering coefficient (b), average distances (c) and
    node betweenness centrality (d)] as a function of average degree for
    BA networks of size $N=10000$.}
  \label{fig:time_deg}
\end{figure*}

Figure~\ref{fig:time_deg}(a) depicts the execution time, in terms of
the average degree, required to calculate the average degrees of BA
networks with size $N = 10000$.

While the speed of the matrix implementations do not depend on the
average degree of the original network, the relative efficiency of the
list representation is dramatic for average degree values up to about
80, decreasing progressively for larger values, until becoming rather
ineffective.  With the matrix representation, the computation of the
average degree involves adding along the rows, which has a constant
cost, therefore becoming independent of the network average degree.
On the contrary, in the case of list representation the average degree
is calculated over a varying number of elements which grow linearly
with the network degree.

\subsubsection{Clustering coefficient}

The dependency of the execution times required for calculation of the
clustering coefficient is shown in Figure~\ref{fig:time_deg}(b), in terms
of several average degrees of BA networks.

The general trends verified in this graph are similar to those
obtained for the networks generation and average degree calculation.
However, the relative advantage of the list implementation is less
marked, becoming less effective for networks with average degrees
larger than 10.

The calculation of the clustering coefficient requires the
identification of the links between the immediate neighbors of the
reference node.  In other words, it is necessary to check the
existence of a link between each pair of nodes $i$ and $j$ connected
to the reference node.  In the case of the matrix representation, this
can be done easily by checking the position $(i,j)$ in the adjacency
matrix.  However, in the case of the list representation, this
requires going through the whole list of nodes that are adjacent to
node $i$ while searching for node $j$.

\subsection{All-pairs distances}

The estimation of the shortest distances in terms of the average
degree is shown in Figure~\ref{fig:time_deg}(c).  The relationship
between the times required for the calculation of these measurements
is similar to the three previous cases, with the difference that the
critical average degree for which the dynamic representation is no
longer the gfastest option is nos between 50 and 60 for the BA model.
This critical degree is certainly dependent of the size of the
network.

\subsection{Betweenness centrality}

Regarding the relationship between the average degree and the
betweeness centrality, shown in Figure~\ref{fig:time_deg}(d), a
relationship similar to that obtained for the two previous cases has
been observed.  However, now we have a higher average degree for which
the dynamic representation becomes worse than the others.  This degree
is dependent of the size of the network, in the case of $N=4000$, this
critical average degree is 100.

\section{Concluding Remarks}

Though the development of more effective algorithms for complex
network generation and characterization can lead to great
computational savings, we have shown that the choice of adequate
network representation can have major impact on the overall execution
time.  More specifically, we compared full and sparse schemes for
representing the connectivity of the networks while generating
networks and calculating several measurement of their topology.  The
sparse representation resulted generally more effective than the full
scheme, with the exception of the cases when the networks have very
large average degree.  We also investigated the effect of having
diverse data types such as byte, integer, float, double and bit.  In
general, the shorter data types led to superior performance as a
consequence of the smaller amount of memory to be accessed.

The obtained results and trends suggest a number of further
investigations.  For instance, it would be interesting to consider
other network models and measurements, as well as to assess the effect
of different types of hardware, compilers and operating systems.

\begin{acknowledgments}
  Luciano da F. Costa thanks CNPq (301303/06-1 and 573583/2008-0) and
  FAPESP (05/00587-5) for sponsorship.
\end{acknowledgments}

\bibliographystyle{unsrt}
\bibliography{netrepr}
 
\end{document}